\documentclass{article}

\usepackage{spconf,amsmath,graphicx}
\usepackage{amsmath,amssymb,amsthm,amsfonts,color,array,url,tikz,wrapfig}
\usepackage{enumitem}
\usepackage[misc,clock,geometry]{ifsym}
\usepackage{booktabs}

\usepackage{placeins}

\usetikzlibrary{positioning}

\title{Nix-TTS: Lightweight and End-to-End Text-to-Speech \\ via Module-wise Distillation}
\name{
     Rendi Chevi$^1$,
     Radityo Eko Prasojo$^{1,2}$,
     Alham Fikri Aji$^{3*}$\thanks{$^*$Work done prior to joining Amazon.},
     Andros Tjandra$^4$,
     Sakriani Sakti$^5$
}
\address{
   $^1$Kata.ai Research Team (ID),
   $^2$Universitas Indonesia (ID),
   $^3$Amazon (UK),
   $^4$Meta AI (USA),\\
   $^5$Japan Advanced Institute of Science and Technology (JP)
}

\begin{document}
\copyrightnotice{978-1-6654-7189-3/22/\$31.00~\copyright2023 IEEE}

\setlength{\abovedisplayskip}{2pt}
\setlength{\belowdisplayskip}{3pt}
%
\maketitle
%
\begin{abstract}
    Several solutions for lightweight TTS have shown promising results. Still, they either rely on a hand-crafted design that reaches non-optimum size or use a neural architecture search but often suffer training costs. We present Nix-TTS, a lightweight TTS achieved via knowledge distillation to a high-quality yet large-sized, non-autoregressive, and end-to-end (vocoder-free) TTS teacher model. Specifically, we offer module-wise distillation, enabling flexible and independent distillation to the encoder and decoder module. The resulting Nix-TTS inherited the advantageous properties of being non-autoregressive and end-to-end from the teacher, yet significantly smaller in size, with only 5.23M parameters or up to 89.34\% reduction of the teacher model; it also achieves over 3.04$\times$ and 8.36$\times$ inference speedup on Intel-i7 CPU and Raspberry Pi 3B respectively and still retains a fair voice naturalness and intelligibility compared to the teacher model. We provide pretrained models and audio samples of Nix-TTS\footnote{\url{https://github.com/rendchevi/nix-tts}}.
\end{abstract}

\section{Introduction}
\label{sec:intro}

Synthetic voices generated by several recent neural TTS models~\cite{ren2020fastspeech, donahue2020end, kim2020glow, kim2021conditional} have been able to achieve high naturalness and intelligibility compared to the real recordings it was trained on. However, most of the models are known to be quite large in size and suffer from slow CPU inference. These limitations hinder the possibility of deploying a real-time, highly natural, and intelligible voice-based interface in low-cost and resource-constrained settings. Although there have been several embedded devices with neural accelerator such as NVIDIA Jetson Nano or Google Coral Dev Board to mitigate said problems, the cost and its availability are still varies. The most challenging is to deploy a TTS in a low-cost CPU-bound devices. In such settings, there are needs for neural TTS models to be inherently lightweight and fast, yet have a good naturalness and intelligibility.

Recent works on lightweight neural TTS models have shown promising results in fulfilling those needs. Several lightweight neural acoustic models design are proposed in~\cite{vainer2020speedyspeech, lee2021bidirectional, beliaev2021talknet}. The size of these models ranges from $4.3$M to $15$M, more than a half smaller compared to established high-quality TTS like~\cite{tacotron2, ren2020fastspeech} which is around $30$M in size. However, these models mainly focus on text-to-Mel, requiring additional neural vocoder to synthesize the waveforms, which variably inflates the model size depending on the chosen vocoder models. The work by ~\cite{nguyen2021litetts} then proposed an end-to-end design of a lightweight TTS model eliminating the need for a neural vocoder. But unfortunately, it could only achieve the final size of above $10$M (about $13$M) parameters which is not yet optimal.

Another solution is to utilize neural compression methods that do not necessarily require manually designing the model and can reduce the size even further. To date, ~\cite{luo2021lightspeech} proposed applying neural architecture search to automatically design an acoustic model; the discovered model is only $1.8$M in size while retaining good naturalness. However, the training cost is arguably high, as it requires sorting through all the possible architecture in the defined search space, making the approach not always feasible for researchers with limited computing resources. Another work by~\cite{lai2021interplay} shows that a neural acoustic model is highly prunable. Even after pruning $90\%$ of its original model (achieving around $3.0$M parameters in size), the pruned model still retains perceivable intelligibility. However, the naturalness of the generated voices is degraded.



As can be seen, the existing works still either rely on a hand-crafted design that reaches non-optimum size or use a neural architecture search but often suffer training costs. Therefore, a novel solution that can achieve a very small-size model without heavy training costs while keeping the performance is necessary.

In this work, we present Nix-TTS, a lightweight TTS achieved via knowledge distillation (KD)~\cite{hinton2015distilling} to a high-quality yet large-sized, non-autoregressive, and end-to-end (vocoder-free) TTS teacher model. However, in contrast with previous work~\cite{vainer2020speedyspeech} that performed KD by only distilled the teacher network’s duration or the work by ~\cite{oord2018parallel, ping2018clarinet} that only utilized KD by compressing neural vocoders, our proposed approach was performed on an end-to-end TTS model. Specifically, we offer a novel module-wise distillation, enabling flexible and independent distillation to the encoder and decoder module. The overall process does not suffer from the training cost and the resulting Nix-TTS:
(1) inherits the teacher properties of being a non-autoregressive and end-to-end framework without the need for an additional vocoder,
(2) achieves a significantly smaller size and inference speedup,
(3) while keeping a fair voice naturalness and intelligibility compared to the teacher model. In addition, we tested our model in CPU-bound devices, a single-thread Intel i7 CPU and Raspberry Pi Model 3B, without any type of neural acceleration. 

\section{Method}
\label{sec:method}
\subsection{Problem Formulation}
\label{sec:problem-formulation}
Let $\mathcal{F}(\cdot ; \omega)$ be an end-to-end neural TTS model. 
We restrict the term ``end-to-end'' TTS model as a model that can generate speech data $x$ in raw waveform $x_{w}$ from text $c$ directly without the need of an external vocoder. Though the architecture of end-to-end TTS varies, during inference, $\mathcal{F}$ can typically be composed of encoder $\mathcal{E}$ and decoder $\mathcal{D}$ as follows:
\vspace{0.05cm}
\begin{equation}
\mathcal{F} = \mathcal{D} \circ \mathcal{E}, \quad x_w = \mathcal{F}(c) = \mathcal{D}(\mathcal{E}(c)), \quad z = \mathcal{E}(c),
\label{eq:e2e-tts-model}
\end{equation} 
where $\mathcal{E}$ encodes $c$ to latent representation $z$, then $\mathcal{D}$ decodes $z$ into $x_w$. Depending on the model, $z$ can be deterministic as in~\cite{ren2020fastspeech, tacotron2} or generative as in~\cite{kim2020glow, kim2021conditional} such that $z \sim \mathcal{N}(\mu, \sigma)$ or any other probability distribution.

Then, in a KD setting, let $\mathcal{F}_{t}$ and $\mathcal{F}_{s}$ be a teacher and a student end-to-end TTS model, respectively, following formulation (\ref{eq:e2e-tts-model}) above. And, let $\{z,x_w\}$ and $\{\hat{z},\hat{x}_w\}$ be the outputs generated by the teacher and student models, respectively.  Given applicable loss functions $\mathcal{L}_\mathcal{E}$ and $\mathcal{L}_\mathcal{D}$, our goal is to design and train $\mathcal{F}_{s}$, so that $\mathcal{E}_{s}$ and $\mathcal{D}_{s}$ satisfy
\vspace{0.1cm}
\begin{equation}
    \text{argmin}_{\hat{z}}\:\mathcal{L}_\mathcal{E}(z,\hat{z}),\;
    \text{argmin}_{\hat{x}_w}\:\mathcal{L}_\mathcal{D}(x_w,\hat{x}_w),\;
    |\omega_s| \ll |\omega_t|,
\end{equation} \vspace{0.1cm}
that is, $\mathcal{F}_{s}$ generates $\hat{z}$ (via $\mathcal{E}_{s}$) and $\hat{x}_w$ close to its counterparts of $\mathcal{F}_{t}$, all the while being significantly smaller.
 
\subsection{End-to-end TTS Teacher}
\label{sssec:e2e-teacher}
We choose VITS~\cite{kim2021conditional} to be our teacher model $\mathcal{F}_t$. It is one of the few fully end-to-end, non-autoregressive, and high-fidelity TTS models. VITS has proven to beat two of the popular non end-to-end TTS models, Tacotron-2~\cite{tacotron2} and Glow-TTS~\cite{kim2020glow} in terms of speech quality while being around the same size ($\sim$30M parameters). Besides the speech quality, the large size and the model complexity make VITS a suitable teacher model for our KD.
\subsubsection{Model Formulation}
\label{sec:model-formulation}
VITS is formulated as a conditional variational autoencoder (cVAE) augmented with normalizing flow and generative adversarial network (GAN). Following the cVAE framework proposed in VITS, let $q_{\theta}(z|x)$ and $p_{\phi}(x|z)$ be the posterior and data distributions respectively, parameterized by neural network's parameters $\theta$ and $\phi$, where $x$ is the speech data variables and $z$ is the latent variables. The prior of which $z$ will be sampled from is defined as $p_{\psi}(z|c)$, where the latents are conditioned on input texts $c$, the prior distribution is parameterized by a neural network's parameters $\psi$. VITS aims to learn the underlying distribution of $x$ given $c$, denoted as $p(x|c)$ by maximizing its evidence lower bound (ELBO).
\vspace{0.1cm}
\begin{multline}
\log p(x|c) \geq \\ \underbrace{-D_{kl}(q_{\theta}(z|x)||p_{\psi}(z|c))}_{\text{KL term}} + \underbrace{\mathbb{E}_{q_{\theta}(z|x)}[\log p_{\phi}(x|z)]}_{\text{Reconstruction term}}.
\end{multline}

In practice, the reconstruction term is implemented as L1 loss between the ground truth and predicted speech data in mel-spectrogram form, $x_m$ that is:
\vspace{0.1cm}
\begin{equation}
\mathcal{L}_{recon} = ||x_{m} - \hat{x}_{m}||_{1},
\label{eq:vits-recon-term}
\end{equation}
where $\hat{x}_{m} \sim p_{\phi}(x|z)$.

Architecture-wise, VITS can be broken down into 3 modules, each encoding the distributions $q_{\theta}(z|x)$, $p_{\phi}(x|z)$, and $p_{\psi}(z|c)$. We describe the role of each module in encoding and decoding the relevant features to be potentially distilled below.

\begin{description}[leftmargin=0.0cm, labelsep=0.25cm]
   \vspace{0.05cm}
   \item[Posterior Encoder] The module consists of non-causal WaveNet residual blocks~\cite{prenger2019waveglow} to encode $x$ in linear spectrogram form, $x_{s}$, into $\{\mu_{q}, \sigma_{q}\}$, the parameters of $q_{\theta}(z|x) = \mathcal{N}(\mu_{q}, \sigma_{q})$. The module infers the latent samples $z_{q} \sim \mathcal{N}(\mu_{q}, \sigma_{q})$ which will then be passed to the Decoder to be reconstructed back to $x$ in raw waveform $x_{w}$.
   \vspace{0.05cm}
   \item[Prior Encoder] The module consists of Transformer encoder blocks~\cite{vaswani2017attention} and a normalizing flow $f$, with affine coupling layers~\cite{dinh2016density}. It encodes $c$ into $\{\mu_{p}, \sigma_{p}\}$, the parameters of $p_{\psi}(z|c) = \mathcal{N}(\mu_{p}, \sigma_{p})$, and prior latent samples, $z_{p} = f(z_{q})$. The alignment between $\{\mu_{p}, \sigma_{p}\}$ and $z_{p}$ is then performed with Monotonic Alignment Search (MAS)~\cite{kim2020glow}. The aligned prior's parameters is denoted as $\{\mu_{p}', \sigma_{p}'\}$. During inference, the network directly infers $z_{q}$ from $f^{-1}(\mu_{p}', \sigma_{p}')$, without needing $x_{s}$.
   \vspace{0.05cm}
   \item[Decoder] The module follows HiFi-GAN V1 generator architecture~\cite{kong2020hifi}. It learns to reconstruct $z_{q}$ into $x_{w}$ with the help of multi-period discriminator~\cite{kim2021conditional} in an adversarial fashion. 
\end{description}

\subsubsection{Available Knowledge to be Distilled}
\label{sec:avail-know}
Assuming the teacher VITS is already trained, we can frame the model in encoder-decoder structure (Section~\ref{sec:problem-formulation}). The natural configuration would be that the Prior Encoder serves as $\mathcal{E}_{t}$ that models the latent distribution $q_{\theta}(z|x)$, and the Decoder serves as $\mathcal{D}_{t}$ that decodes $x_{w}$ from latent samples $z_{q} \sim q_{\theta}(z|x)$. 

However, the Posterior Encoder also encodes the same latent samples as the Prior Encoder, making both encoders suitable as $\mathcal{E}_{t}$. This means that the student $\mathcal{E}_{s}$ can actually be distilled from either, although the Prior Encoder is more complex due to the presence of $f$ that only provides the stochastic samples of $q_{\theta}(z|x)$, instead of its deterministic parameters. Thus, we opt to distil $q_{\theta}(z|x)$ from the Posterior Encoder as it is arguably easier given the right student model.

\subsection{End-to-end TTS Student}
\label{sec:nix-tts}
\begin{figure}[t]
  \centering
  \includegraphics[width=0.43\textwidth]{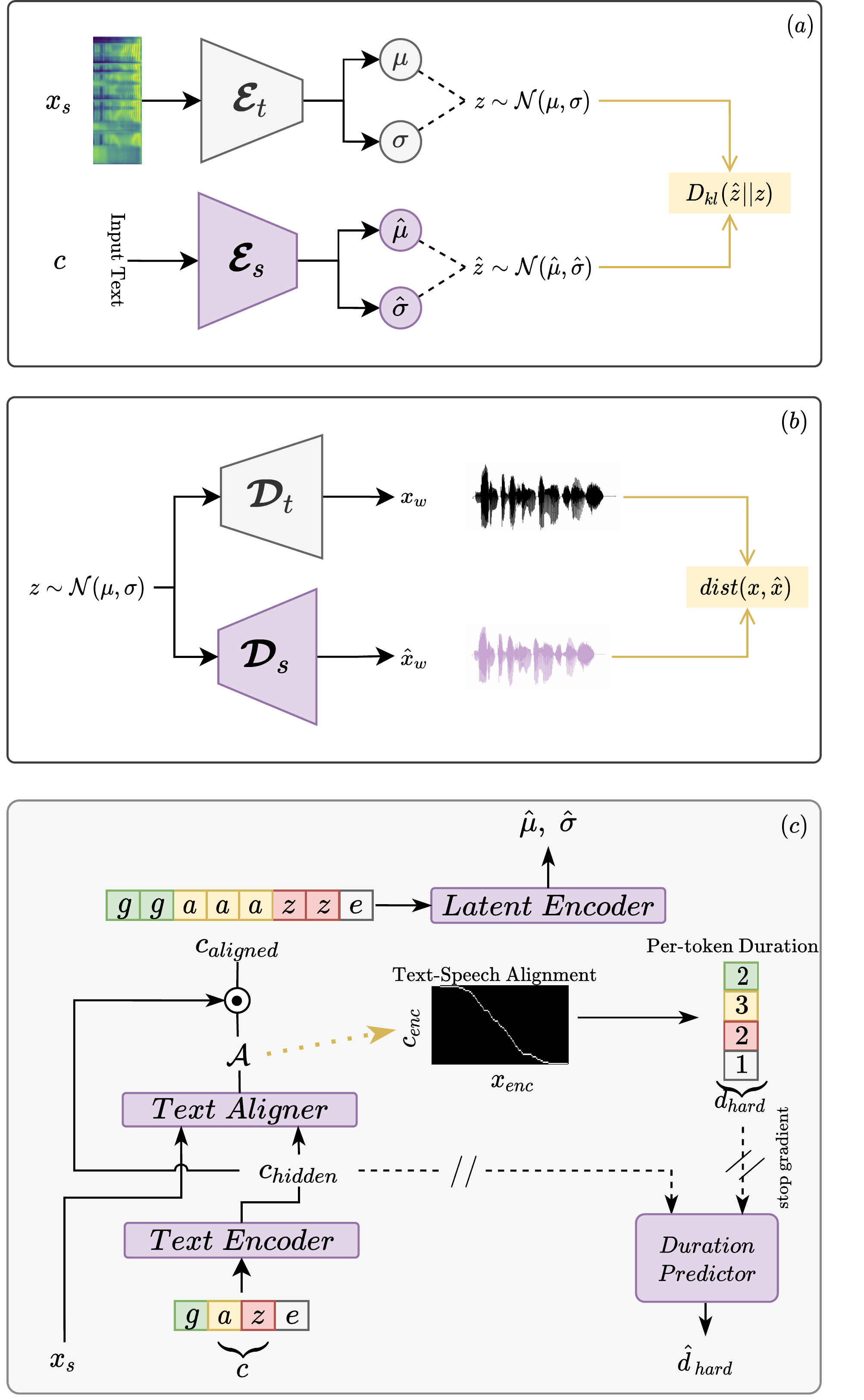}
    \vspace{-0.2cm}
    \caption{Illustration of Nix-TTS Module-wise Distillation. (a) Encoder distillation, the student encoder distills the teacher's speech latent distribution given input texts, it learns the text-speech alignment by their own as described in Section~\ref{sec:student-encoder}; (b) Decoder distillation, student encoder distills output audio waveform given the teacher's speech latent samples; (c) Nix-TTS encoder architecture that transforms input texts into teacher's speech latent parameters.}
  \label{fig:nix_diag}
\end{figure}

This section describes our proposed model, Nix-TTS. Nix-TTS serves as the end-to-end TTS student model $\mathcal{F}_s$ to be distilled from VITS as the end-to-end TTS teacher $\mathcal{F}_t$. 

\subsubsection{Encoder Architecture}
\label{sec:student-encoder}
Nix-TTS’s encoder, $\mathcal{E}_{s}$, aims to model $q_{\theta}(z|x) = \mathcal{N}(\mu_{q}, \sigma_{q})$, by predicting its parameters, $\{\mu_{q}, \sigma_{q}\}$, as shown in Fig.~\ref{fig:nix_diag}. Because $q_{\theta}(z|x)$ is conditioned on $x_{s}$ instead of $c$ which has different modality and dimensions, we need to encode $c$ so that it is meaningfully aligned with the corresponding $x_{s}$. To achieve this, we compose $\mathcal{E}_{s}$ out of 4 main modules: (1) Text Encoder, (2) Text Aligner, (3) Duration Predictor, and (4) Latent Encoder. We detail each of those modules below.

\begin{description}[leftmargin=0.0cm, labelsep=0.25cm]
   \vspace{0.05cm}
   \item[Text Encoder] The module encodes $c$ into text hidden representation $c_{hidden}$. It feeds $c$ through an embedding layer followed by an absolute positional encoding~\cite{vaswani2017attention} and stacks of dilated residual 1D convolutions blocks. Each convolution block is followed by a SiLU (Sigmoid Linear Unit) activation~\cite{swish2017} and a layer normalization~\cite{vaswani2017attention}. We apply alternating dilation rates on each block~\cite{vainer2020speedyspeech} to increase the network's receptive field.
   \vspace{0.05cm}
   \item[Text Aligner] To learn the $c$ and $x_{s}$ alignment, we adopt the framework proposed by~\cite{shih2021radtts, badlani2021one}. The module first encodes $c_{hidden}$ and $x_{s}$ into $c_{enc}$ and $x_{enc}$ with a series of convolutions. The ``soft'' alignment $\mathcal{A}_{soft}$ is then generated by taking the normalized pairwise affinity~\cite{badlani2021one} between the two. As a non-autoregressive TTS requires ``hard'' durations to be defined per token, MAS~\cite{kim2020glow} is applied to $\mathcal{A}_{soft}$ to generate $\mathcal{A}_{hard}$. The aligned text representation $c_{aligned}$ can then be generated by applying batch matrix-matrix product between $c_{hidden}$ and $\mathcal{A}_{hard}$. The generated alignments are defined as follows:
   \vspace{0.05cm}
    \begin{equation}
    \mathcal{A}_{soft} = \text{softmax}(\sum_{j=1}^{J}\sum_{k=1}^{K} (c_{enc}^{j} - x_{enc}^{k})^{2})
    \end{equation}
    \begin{equation}
    \mathcal{A}_{hard} = \text{MAS}(\mathcal{A}_{soft}).
    \end{equation}
    \vspace{0.05cm}
   \item[Duration Predictor] The module serves to predict $\mathcal{A}_{hard}$ without the need of $x_{s}$ during inference.
    The module composed of stacks of 1D convolutions to predict the per-token durations $d_{hard}$ extracted from $\mathcal{A}_{hard}$ (Eq. 7). It is formulated as a regressor to predict $d_{hard}$ given $c_{hidden}$.
    \vspace{0.1cm}
    \begin{equation}
    d^{j}_{hard} = \sum_{k=1}^{K} \mathcal{A}_{hard}^{k, j}.
    \end{equation}
    \vspace{0.05cm}
   \item[Latent Encoder] The latent encoder follows the same architecture as the text encoder without the embedding layer. The parameters $\{\mu_{q}, \sigma_{q}\}$ is generated by projecting the output of the latent encoder with a single perceptron layer.
\end{description}

\subsubsection{Decoder Architecture}
\label{sec:student-decoder}
Nix-TTS’s decoder, $\mathcal{D}_{s}$, aims to model the distribution $p_{\phi}(x|z)$. It takes the latent variables $z_{q} \sim \mathcal{N}(\mu_{q}, \sigma_{q})$ as inputs and decodes the corresponding raw waveform $x_{w}$.

Unlike $\mathcal{E}_{s}$ which differs highly from $\mathcal{E}_{t}$, $\mathcal{D}_{s}$ roughly follows the same architecture as $\mathcal{D}_{t}$ but with significantly fewer parameters. The decoder follows Hifi-GAN's generator architecture which consists of a series of transposed convolutions and multi-receptive fusion module~\cite{kong2020hifi}. During training, $\mathcal{D}_{s}$ is equipped with a discriminator $\mathcal{C}_{s}$, which follows the teacher's multi-period discriminator architecture~\cite{kim2021conditional}.

To significantly reduce the number of parameters of $\mathcal{D}_{s}$. We replace the vanilla convolutions used in the original architecture with depthwise-separable convolutions~\cite{chollet2017xception} and halve the number of feature maps dimensions. We also reduce the number of parameters of $\mathcal{C}_{s}$ to not overpower the $D_{s}$ reduced capacity. This results in 93\% and 63\% size reduction in $\mathcal{D}_{s}$ and $\mathcal{C}_{s}$ respectively, making the training and inference much faster.

\subsection{Module-wise Distillation}
\label{sec:module-wise}
We distill Nix-TTS in a module-wise fashion, enabling flexible and independent distillation to the encoder and decoder module, as illustrated in Fig.~\ref{fig:nix_diag}. This section describes the training objectives for each encoder and decoder module.

\subsubsection{Encoder Distillation}
The main tasks of $\mathcal{E}_{s}$ are to learn the alignment between $c$ and $x_{s}$ and to model $q_{\theta}(z|x)$ by predicting its parameters. Our alignment objective follows~\cite{shih2021radtts, badlani2021one}, which maximizes the likelihood of $c_{hidden}$ given $x_{s}$ as represented in $\mathcal{A}_{soft}$ using forward-sum algorithm and forcing the match between $\mathcal{A}_{soft}$ and $\mathcal{A}_{hard}$ by minimizing their KL-divergence. The two losses are respectively denoted as $\mathcal{L}_{ForwardSum}$ and $\mathcal{L}_{bin}$ as in~\cite{badlani2021one}.

For modeling $q_{\theta}(z|x)$, we minimize the KL-divergence between $\mathcal{N}(\hat{\mu}_{q},\hat{\sigma}_{q})$ and $q_{\theta}(z|x)$. Because we know both distributions are gaussians, we can minimize the closed-form KL-divergence between two gaussians as:
\vspace{0.05cm}
\begin{equation}
\mathcal{L}_{kl} = \mathbb{E}_{\mu, \sigma}[-\frac{1}{2} + log\frac{\sigma_{q}}{\hat{\sigma}_{q}} + \frac{\hat{\sigma_{q}}^{2} + (\hat{\mu_{q}} - \mu_{q})^{2}}{2\sigma_{q}^{2}}].
\end{equation}

\noindent The final encoder objectives are summarized as follows:
\vspace{0.05cm}
\begin{equation}
\mathcal{L}_{\mathcal{E}} = \mathcal{L}_{ForwardSum} + \mathcal{L}_{bin} + \mathcal{L}_{kl}.
\end{equation}

\subsubsection{Decoder Distillation}
The main task of $\mathcal{D}_{s}$ is to generate $\hat{x_{w}}$ that sounds as similar as possible to $x_{w}$. This is achieved with least-square adversarial training~\cite{mao2017least} (Eq.~\ref{eq:decoder-adv-1} and~\ref{eq:decoder-adv-2}), feature-matching loss~\cite{larsen2016autoencoding} (Eq.~\ref{eq:decoder-fmaps}), and mel-spectrogram reconstruction loss (Eq.~\ref{eq:vits-recon-term}, that is, the same as VITS' reconstruction term). 
\vspace{0.05cm}
\begin{equation}
\mathcal{L}_{adv, disc} = \mathbb{E}_{x_{w}}[\mathcal{C}_{s}(x_{w}) - 1)^{2} + \mathcal{C}_{s}(\hat{x}_{w})^{2}]
\label{eq:decoder-adv-1}
\end{equation}
\begin{equation}
\mathcal{L}_{adv, gen} = \mathbb{E}_{\hat{x}_{w}}[\mathcal{C}_{s}(\hat{x}_{w}) - 1)^{2}]
\label{eq:decoder-adv-2}
\end{equation}
\begin{equation}
\mathcal{L}_{fmatch} = \sum_{l=1}^{l=L} \frac{1}{n_{l}}||\mathcal{C}_{s}^{l}(x_{w}) - \mathcal{C}_{s}^{l}(\hat{x}_{w})||_{1}.
\label{eq:decoder-fmaps}
\end{equation}
Note that $C_{s}^{l}$ is the feature maps in the discriminator's $l^\text{th}$ layer, $n_{l}$ is the number of feature maps in $l^\text{th}$ layer, and $L$ is the number of layers of $\mathcal{C}_{s}$.

We also augment our decoder objective with generalized energy distance (GED) loss~\cite{gritsenko2020spectral} in order to accelerates the training convergence and improves the audio quality, defined as
\begin{equation}
\mathcal{L}_{ged} = \mathbb{E}_{x_{w}}[2d_{spec}(x_w, \hat{x}_{w}^{a}) - d_{spec}(\hat{x}_{w}^{a}, \hat{x}_{w}^{b})],
\end{equation}
where $d_{spec}(.)$ is a multi-scale spectrogram distance as seen in~\cite{engel2020ddsp} whereas $\hat{x}_{w}^{a}$ and $\hat{x}_{w}^{b}$ are audio generated from $\mathcal{D}_{s}$ conditioned on the same $\hat{\mu}_{q}, \hat{\sigma}_{q}$ but different noise samples drawn from $\mathcal{N}(0, 1)$. The final decoder objective is then computed as:
\begin{equation}
\mathcal{L}_{\mathcal{D}} = \mathcal{L}_{adv, disc} + \mathcal{L}_{adv, gen} + \mathcal{L}_{fmatch} + \mathcal{L}_{recon} + \mathcal{L}_{ged}.
\end{equation}

\section{Experiments}

\begin{description}[leftmargin=0.0cm, labelsep=0.25cm]
   \item[Dataset] 
We use the LJSpeech dataset~\cite{ljspeech17} consisting of 13,100 utterances (about 24 hours of speech) with a 22,050 Hz sampling rate. We split the dataset into 12,500 utterances, 100 utterances, and 500 utterances for training, validation and test set, respectively, following VITS's setup\footnote{\label{vits-repo}\url{https://github.com/jaywalnut310/vits}}.
\item[Teacher Configuration]
We use the VITS model released by the authors~\cite{kim2021conditional} as the teacher, which was trained on the LJSpeech training dataset. For the spectral data processing (linear and Mel spectrogram extraction), we use a filter and window length of 1024, hop size of 256, and Mel-channels of 80. We extracted the teacher's features described in Section~\ref{sec:avail-know} pre-distillation with the mentioned dataset and data processing protocol.
   \item[Student Configuration] The student model follows the architecture in Section~\ref{sec:nix-tts}. We use kernel size of 5, hidden size of 192, and alternating dilation rates of $\{1, 2, 4\}$ for the Text and Latent Encoder. We adopt the implementation of~\cite{lee_comp_tts} for the Text Aligner with the hyperparameters described in~\cite{shih2021radtts}. For the Duration Predictor, we use two 1D convolution layers with kernel size of 3 and hidden size of 192. For the Decoder, we use the modified HiFi-GAN v1 following Section~\ref{sec:student-decoder} with upsample channel of 256.
   \item[Training Configuration] The proposed Nix-TTS is trained following Section~\ref{sec:module-wise} with the extracted teacher's features. We used AdamW optimizer~\cite{loshchilov2017decoupled} with an initial LR of $2 \times 10^{-4}$, $\beta_{1} = 0.8$, $\beta_{2} = 0.99$, and $\lambda = 0.1$. We also apply LR decay exponentially by a factor of $0.999875$ per epoch. All models are trained on a single RTX 3090 GPU with a total batch size of 32.
    \item[Evaluation Configuration] We evaluate the models from various factors: speech intelligibility, naturalness, model parameter size, model speedup, and complexity. Aside from comparing with the teacher VITS, we also compare with existing lightweight TTS models that are popular and publicly available: BVAE-TTS~\cite{lee2021bidirectional} and SpeedySpeech~\cite{vainer2020speedyspeech}. However, as BVAE-TTS and SpeedySpeech are only text-to-Mel acoustic models, we use HifiGAN~\cite{kong2020hifi} as their vocoders, specifically HifiGAN-v1-Universal publicly available vocoder\footnote{https://github.com/jik876/hifi-gan}.
\end{description}

\section{Results}
\subsection{Speech Synthesis Quality}
\subsubsection{Naturalness}

The speech quality of naturalness of the teacher VITS in LJ Speech datasets has been discussed in~\cite{kim2021conditional} where it reached 4.43 MOS that successfully beat Tacotron-2~\cite{tacotron2} and Glow-TTS~\cite{kim2020glow}. Here, we first investigate how much the student Nix-TTS can retain the speech quality of the teacher VITS. We conduct CMOS (Comparison Mean Opinion Score) evaluation following the well-known LightSpeech~\cite{luo2021lightspeech}, which shares a similar goal as ours with a different compression approach.
CMOS value ranges from [-3, +3], where more negative score indicates the audio is less preferable and the positive score indicates the audio is more preferable. Our results (Table~\ref{tab:cmos}) from 23 subjects show that our Nix-TTS model is only slightly less preferred than the teacher model, while being significantly smaller in size by 82\%. The CMOS result demonstrates a good model size vs. speech quality trade-off.

\begin{table}[h]
  \caption{Comparison Mean Opinion Score}
  \vspace{0.25cm}
  \label{tab:cmos}
  \footnotesize
  \centering
  \begin{tabular}{lrr}
    \toprule
    \textbf{Model} & $\downarrow$\textbf{Params} & $\uparrow$\textbf{CMOS} [-3, +3] \\
    \midrule
    VITS~\cite{kim2021conditional} & $29.08$M & $0.00$ \\
    \midrule
    Proposed Nix-TTS & $5.23$M & $-0.27$ \\
    \bottomrule
  \end{tabular}
  \vspace{-0.4cm}
\end{table}

\subsubsection{Intelligibility}
Next, we evaluate the intelligibility of our model and the baseline with an off-the-shelf ASR (Automatic Speech Recognition) model, inspired from~\cite{nguyen2021litetts}. We choose pretrained Wav2Vec2.0 English ASR model~\cite{baevski2020wav2vec} publicly available in Hugging Face\footnote{https://huggingface.co/facebook/wav2vec2-large-960h}. To conduct the evaluation, we first generate audio samples for all the TTS models with the texts from the test split of LJSpeech as seen in~\cite{kim2021conditional}. We then feed the generated audio samples to the ASR model and infer the predicted text. Finally, we evaluate the PER (Phoneme Error Rate) of the predicted texts against the ground-truth texts.

\begin{table}[h]
  \caption{Phoneme Error Rate (PER)}
  \vspace{0.25cm}
  \label{tab:per}
  \footnotesize
  \centering
  \begin{tabular}{lrrr}
    \toprule
    \textbf{Model} & $\downarrow$\textbf{Params} & $\downarrow$\textbf{PER (\%)}\\
    \midrule
    VITS~\cite{kim2021conditional} & $29.08$M & $1.51$ \\
    \midrule
    BVAE-TTS+HifiGAN ~\cite{vainer2020speedyspeech} & $15.99 (+13.93)$M & $1.85$ \\
    SpeedySpeech+HifiGAN ~\cite{lee2021bidirectional} & $4.52 (+13.93)$M & $3.00$ \\
    Proposed Nix-TTS & $5.23$M & $2.07$ \\
    \bottomrule
  \end{tabular}
\end{table}

From the result, Nix-TTS' PER score degraded only by 0.5\% from the teacher, achieving the second best PER among the baseline models. Our proposed model able to have lower PER than SpeedySpeech by a fair margin. While the PER of BVAE-TTS is slightly lower than Nix-TTS by 0.2\%, the model size of Nix-TTS is much smaller than BVAE-TTS. The result demonstrates our proposed model achieves a good model size vs. speech intelligibility trade-off.




\subsection{Model Speedup and Complexity}
We compare the models speedup and complexity in terms of RTF (Real Time Factor) and number of parameters in 2 scenarios: on a single thread Intel-i7 CPU and on a Raspberry Pi Model B. The latter is to simulate a real-world scenario of deploying a TTS in low-cost and resource-constrained settings. For the latter scenario, we transform the model into an optimized ONNX version\footnote{https://github.com/onnx/onnx} to reduce library dependencies which typically not available in an embedded device.

\newcolumntype{L}[1]{>{\arraybackslash}p{#1}}
\begin{table}[h]
    \caption{Model Speedup and Complexity}
    \vspace{0.25cm}
    \centering
    \footnotesize
    \setlength\tabcolsep{2.5pt}
    \begin{tabular}{lcccc}
    \toprule
    \multicolumn{5}{c}{\textbf{Intel Core i7 CPU @ 1.10GHz (Single Thread)}} \\
    \toprule
    \textbf{Model} & $\quad\downarrow$\textbf{Params}$\quad$ & $\uparrow$\textbf{CompRat} & $\quad\downarrow$\textbf{RTF}$\quad$ & $\uparrow$\textbf{SpdRat} \\
    \midrule
    VITS~\cite{kim2021conditional} & $29.08$M & $-$ & $0.484$ & $-$ \\
    \midrule
    BVAE-TTS & $29.92$M & $0\%$ & $0.383$ & $1.26\times$ \\
    +HifiGAN~\cite{lee2021bidirectional}& &&&\\
    SpeedySpeech & $18.45$M & $36.55\%$ & $0.327$ & $1.48\times$ \\
    +HifiGAN~\cite{vainer2020speedyspeech} & &&&\\
    Proposed Nix-TTS & $5.23$M & $89.34\%$ & $0.159$ & $3.04\times$ \\
    \toprule
    \multicolumn{5}{c}{\textbf{Raspberry Pi Model 3B}} \\
    \toprule
    \textbf{Model} & $\downarrow$\textbf{Storage} & $\uparrow$\textbf{CompRat} & $\downarrow$\textbf{RTF} & $\uparrow$\textbf{SpdRat} \\
    \midrule
    VITS$^{\dagger}$~\cite{kim2021conditional}  & $113.5$MB & $-$ & $16.50$ & $-$ \\
    \midrule
    BVAE-TTS &$123.1$MB & $0\%$ & $16.11$ & $1.02\times$ \\
    +HifiGAN$^{\dagger}$~\cite{lee2021bidirectional}  & &&&\\
    SpeedySpeech &$73.1$MB & $35.59\%$ & $15.93$ & $1.03\times$ \\
    +HifiGAN$^{\dagger}$~\cite{vainer2020speedyspeech}  & &&&\\
    Proposed Nix-TTS$^{\dagger}$  & $21.2$MB & $80.88\%$ & $1.974$ & $8.36\times$ \\
    \bottomrule
    \multicolumn{5}{L{0.95\columnwidth}}{$\dagger$ denotes the ONNX version of the model. \textbf{CompRat} is Compression Ratio relative to the teacher VITS. \textbf{RTF} is the Real Time Factor value. \textbf{SpdRat} is Speedup Ratio relative to the teacher VITS.
    }
  \end{tabular}
  \label{tab:speedcomp}
\end{table}

\FloatBarrier

The results are summarized in Table~\ref{tab:speedcomp}. On a single thread Intel Core i7 CPU @ 1.10GHz, Nix-TTS achieves the largest speedup with 3.04x faster generation than the teacher VITS, alongside with 89.34\% reduction of model size. On Raspberry Pi Model B, where all the models are converted into the same ONNX format, the speedup is even more pronounced with 8.36x speedup alongside 81.32\% reduction of storage size. Moreover, investigating the speedup module-wise, the inference speed in seconds for student and teacher's encoders are ~0.01 sec. vs. ~0.11 sec., while the decoders are ~0.16 sec. vs. ~1.06 sec. We mainly attribute the significant speedup to our use of the more computationally efficient depth-wise separable convolutions instead of using self-attentions.

The model size reduction and speedup ratio of Nix-TTS against the other baseline models linearly follows as the other models' compression and speedup ratio are below Nix-TTS. In addition, due to the proposed model already small enough to be run on a device without neural acceleration, we could assume that our model would perform much better in hardware with a neural accelerator such as NVIDIA Jetson Nano, Google Coral Dev Board, or other consumer-grade GPUs. The results demonstrate Nix-TTS's advantage as a lightweight TTS in low-cost and resource-constrained settings.

\section{Conclusion}
We proposed Nix-TTS, a lightweight and end-to-end TTS model achieved by performing KD to a high-quality yet large-sized, non-autoregressive, and end-to-end TTS teacher model. To distill the non-autoregressive TTS teacher, we present a module-wise distillation technique enabling flexible and independent distillation to the encoder and decoder module. Our proposed model retains good naturalness and intelligibility vs. model size trade-off compared to the teacher and other existing lightweight TTS models. All the while being as small as $5.23$M parameters in size and achieves 3.04× and 8.36× speedup on Intel-i7 CPU
and Raspberry Pi, making it advantageous for low-cost and resource-constrained settings.


\bibliographystyle{IEEEbib}
\bibliography{main_draft}

\end{document}